\documentclass[twocolumn,a4paper]{revtex4}

\begin{document}

\title{An alternative theoretical approach to describe planetary
systems through a Schr\"{o}dinger-type diffusion equation}

\author{M. de Oliveira Neto$^1$, L. A. Maia$^2$ and S.
Carneiro$^3$}

\affiliation{$^1$Instituto de Qu\'{\i}mica, Universidade de
Bras\'{\i}lia, 70910-900, Bras\'{\i}lia, Brazil\\ $^2$Departamento
de Matem\'{a}tica, Universidade de Bras\'{\i}lia, 70910-900,
Bras\'{\i}lia, Brazil\\ $^3$Instituto de F\'{\i}sica, Universidade
Federal da Bahia, 40210-340, Salvador, BA, Brazil}

\begin{abstract}
In the present work we show that planetary mean distances can be
calculated with the help of a Schr\"odinger-type diffusion
equation. The obtained results are shown to agree with the
observed orbits of all the planets and of the asteroid belt in the
solar system, with only three empty states. Furthermore, the
equation solutions predict a fundamental orbit at $0.05$ AU from
solar-type stars, a result confirmed by recent discoveries. In
contrast to other similar approaches previously presented in the
literature, we take into account the flatness of the solar system,
by considering the flat solutions of the Schr\"odinger-type
equation. The model has just one input parameter, given by the
mean distance of Mercury.
\end{abstract}

\maketitle

\section{Introduction}

In recent years, some authors \cite{Marcal}-\cite{Arp} have
suggested a quantum-like approach to calculate the planetary
orbits in our solar system, which has led to several impressive
results. Besides obtaining the observed orbits of all the planets
and of the asteroid belt, such models have made predictions
subsequently confirmed by observations.

For example, the existence of asteroids orbiting between Uranus
and Neptune, with orbit radii around $24.77$ AU \cite{Marcal}, was
confirmed by recent discoveries \cite{MPC}. Furthermore, the
prediction of two intramercurial orbits with radii around $0.05$
AU and $0.18$ AU \cite{Nottale2,AF1}, absents of our solar system,
has been verified in several extra-solar planetary systems during
the last years \cite{Schneider}.

In spite of these and others positive results, these approaches
seem to be very speculative, for the quantization of macroscopic
systems is something outside the scope of our known physics.
Nevertheless, some possible origins for such effects have been
outlined, some of them on the basis of quite orthodox theories.

A large scale quantization can be based, for instance, on an
extension of the ordinary commutation rules, in order to recover
the equivalence principle in the context of quantum mechanics
\cite{AF1,Greenberger}. The general commutation rules derived in
this way predict two scales of quantization: the usual microscopic
one, and the scale in which the gravitational interactions are
dominant. The presence of quantization at so different scales can
be related as well to self-similarity concepts \cite{SC}. And it
is also characteristic of fractal space-time approaches
\cite{Nottale1}, in which the particle trajectories are
non-differentiable at very short and very large scales.

The concept of non-differentiability of space-time at short scales
has been used by some authors in trying to derive the
Schr\"{o}dinger equation in the context of a classical framework.
For example, in the stochastic mechanics of Nelson \cite{Nelson},
the Schr\"{o}dinger equation is obtained as a classical diffusion
equation, with help of the hypothesis that any particle in the
empty space, under the influence of any interaction field, is also
subject to a universal Brownian motion without viscosity. Under
this hypothesis, a Schr\"{o}dinger-type equation follows from
classical dynamics, with the Planck constant being related to a
diffusion coefficient.

The main problem of this kind of derivation of the Schr\"{o}dinger
equation (as well as of other descriptions based on diffusion
processes or fluid dynamics \cite{Naschie}) is looking for a
convincing physical origin for that universal Brownian motion,
although some possibilities have been outlined, based for instance
on quantum fluctuations on cosmic scale \cite{Sidharth}, or on the
quantum nature of space-time in quantum gravity theories
\cite{Bergia}.

Nevertheless, the important point in Nelson's work is that a
diffusion process can be described in terms of a
Schr\"{o}dinger-type equation. In this context, the possibility of
describing a classical process like the formation of solar system
in terms of quantum mechanics can be seriously considered. The use
of such an approach is also suggested by the chaotic behaviour of
the solar system during its formation and evolution
\cite{Hills}-\cite{SW}, which implies the non-differentiability of
trajectories for large time scales \cite{Nottale3}.

Despite all these conjectures, the physical principles behind the
quantization of large structures cannot be considered completely
understood. In the present work, our aim is not to solve this
problem, but just to explore a little more the results we can
obtain from a quantum-like approach. Our goal is to introduce a
new ingredient not considered up to now: the flatness of the solar
system. For this purpose, we will look for flat solutions of the
Schr\"{o}dinger-type equation, in contrast to the spherical
solutions usually considered in the previous papers.

\section{Preliminary remarks}

Applying a Bohr-like ``quantization" rule to the orbital angular
momentum of a planet in circular orbit around the Sun,
$L=mvr=ng^{*}/2\pi$, and using the Newtonian law for the orbital
velocity, $v^2=GM/r$, one of the authors \cite{Marcal} has derived
the following expression for the radii of the orbits:
\begin{equation}
\label{1} r = \frac{n^2{g^*}^2}{4\pi^2GMm^2},
\end{equation}
where $n$ is an integer, $G$ is the gravitational constant, $M$ is
the solar mass, $m$ is the planet mass and $g^{*}$ stands for a
quantum of action, playing the role of a re-scaled Planck
constant.

In the case $n=1$, equalizing (\ref{1}) to the observed value of
the Mercury orbital radius, using the observed values for $G$ and
$M$, and taking $m=2.1 \times 10^{26}$ kg (the average mass of the
planets of the solar system), we find $g^{*}=3.6 \times 10^{42}$
J.s. This value coincides to a re-scaled quantum of action derived
on the basis of self-similarity considerations \cite{SC}.

Now, equation (\ref{1}) can be written as $r=n^2r_1$, where $r_1$
is the Mercury orbital radius. Taking for $n$ the next integer
values, we can obtain a sequence of values that assures very well
the observed values of orbital radii in our solar system
\cite{Marcal}. For $n=2$ we obtain the orbit of Mars; for $n=3$
the orbit of Camilla (an asteroid located in the outer region of
the asteroid belt between Mars and Jupiter); the values $n=4$ and
$n=5$ give the orbits of Jupiter and Saturn, respectively; for
$n=6$ we have the orbital radius of Chiron (an asteroid between
Saturn and Uranus); and, for $n=7$, $9$ and $10$, the orbits of
Uranus, Neptune and Pluto follow, in this order. The deviations
from the observed values are $19\%$ for Jupiter, $4\%$ for Neptune
and less than $2\%$ in the other cases.

The value $n=8$ predicts an orbit between Uranus and Neptune, with
radius $24.8$ AU. In reference \cite{Marcal}, submitted for
publication in October 1994, this orbit was considered empty.
Surprisingly enough, the value above is in very good agreement
with subsequent discoveries of two asteroids located precisely at
this radial position, named 1993 HA 2 and 1995 DW 2 \cite{MPC}.
These discoveries were published in 1995, that is, one year after
the paper submission, showing the predict power of this kind of
model.

The most important drawback of the model is, as already noted by
the reader, the absence of Venus and Earth in the above sequence.
To include these planets, and also the asteroid Vesta, located in
the inner region of the asteroid belt, it has been taken into
account an {\it ad hoc} second quantum number $n'$, running from
$0$ to $n$ \cite{Marcal}, such that relation (\ref{1}) is
generalized to
\begin{equation} \label{marcal'}
r = \frac{(n^2+n'^2){g^*}^2}{8\pi^2GMm^2}.
\end{equation}
For $n=n'$ we recover the original relation and results.

Another possibility was provided by Agnese and Festa \cite{AF1},
taking for $n=1$, instead of the Mercury radius, the value
$r_1=0.04$ AU, which corresponds to $g^{*}=1.2 \times 10^{42}$
J.s. Now, for $n$ running from $3$ to $6$, we obtain the orbital
radius of Mercury, Venus, Earth and Mars; for $n=8$ we have the
asteroid Ceres; and, for $n=11$, $15$, $21$, $26$ and $30$, the
orbital radius of Jupiter, Saturn, Uranus, Neptune and Pluto
follow, in agreement with the observed values. The problem here,
as we can see, is that we fall into a lot of empty orbital
positions, that is, radius values predicted by the formula
$r=n^2r_1$, but not occupied by any observed body, particularly
for large values of $n$. Nevertheless, the prediction of a
fundamental radius at $0.04$ AU would be shown an important
result.

Another interesting contribution by Agnese and Festa is the
introduction of the re-scaled fine structure constant
$\alpha_g=2\pi GMm/g^{*}c$, where $c$ is the light velocity. In
this way one can rewrite equation (\ref{1}) as
\begin{equation}
\label{2} r = \frac{n^2GM}{\alpha_g^2c^2}.
\end{equation}
Using for $m$ the average mass of planets of our solar system and
for $g^{*}$ the value obtained above, we obtain for $\alpha_g$ the
value used by Agnese and Festa, $\alpha_g^{-1}=2.1 \times 10^3$.

This quantum approach also works for other, recently discovered,
planetary systems, as shown by Nottale \cite{Nottale2,Nottale4}
and also by Agnese and Festa \cite{AF2,AF3}. In Nottale's approach
the periods of planets revolution are quantized as
\begin{equation}
\label{3} T = 2\pi GMn^3/\omega_0^3,
\end{equation}
where $\omega_0=144$ km/s is a characteristic velocity. Using
$T=2\pi r/v$, $v^2=GM/r$ and $r$ given by (\ref{2}), we derive
$\omega_0=\alpha_gc$. Using $\omega_0=144$ km/s, we obtain
$\alpha_g^{-1}=2.1 \times 10^3$, which again is in accordance with
the value obtained by Agnesa and Festa.

Therefore, Nottale also predicts a fundamental radius given by
$r_1\approx 0.04$ AU. Several extra solar planets recently
discovered lie at this distance from their star \cite{Schneider}.
This agreement between a fundamental radius derived years ago and
subsequent observations shows once again the predict power of such
models.

Some of the above results come from a semi-classical treatment of
the problem, based essentially on the application of a Bohr-like
quantization condition. In Nottale's works we can find a more
rigorous approach, based on a Schr\"{o}dinger-type wave equation.
In this way he can accommodate all the planets and the asteroid
belt of our solar system, with only a few empty orbits
\cite{Nottale1}. To achieve this goal he treats the inner and
outer systems separately, with two distinct values of $\omega_0$.

In the next section, we propose to show that flat solutions of the
Schr\"{o}dinger-type equation can furnish important information on
the mean planetary radii. Our aim is to enrich the previous
studies, including a new ingredient not considered up to now: the
flatness of the solar system and of the original disk from which
it was originated.

\section{The flat solutions}


\def\fff{\displaystyle\frac}
\def\cc{\c{c}}
\def\vdois{\vspace*{2mm}}
\def\vcinco{\vspace*{5mm}}
\def\ddd{\displaystyle}
\def\vtres{\vspace*{3mm}}
\def\und{\underline}
\def\raiz{\ddd\sqrt}

In this section we present the mathematical results obtained from
the solution of a Schr\"{o}dinger-type equation involving an
attractive central field. As already mentioned, the orbits of
planets and asteroids will be considered approximately in the same
plane, and the $3$-dimensional equation will be solved taking
$z=0$.

Let us consider a body of mass $m$ moving around another body of
mass $M$, under an attractive central field with potential $V(r)$,
where $r$ is the axial distance in polar coordinates. The
Schr\"{o}dinger-type equation is given then by
\begin{equation}
\label{4} -\fff{{g^*}^2}{2\mu}\left(\fff{\partial^2 \psi}{\partial
r^2} + \fff{1}{r} \fff{\partial \psi }{\partial r} + \fff{1}{r^2}
\fff{\partial^2\psi }{\partial\theta^2}\right)+ V(r)\psi=E\psi.
\end{equation}
The term in parenthesis is $\Delta\psi$, where $\Delta$ is the
Laplace operator in polar coordinates ($r, \theta$), which is
applied to the wavefunction $\psi$ of the body of mass $m$. The
parameter $E$ stands for the total energy of the system; $\mu$ is
the reduced mass $mM/(m+M)$; and $g^*$ is a constant, as
previously described.

As long as the potential $V$ is a function of the radial variable
only, we may look for a solution using separation of variables,
\begin{equation}
\label{5} \psi(r, \theta) = f(r)\Theta(\theta).
\end{equation}
Replacing (\ref{5}) in (\ref{4}) and dividing by $f(r) \Theta
(\theta)$, we get
\begin{eqnarray}
\fff{{g^*}^2}{2\mu} \fff{1}{f(r)} \left[\fff{d^2 f(r)}{d r^2} +
\fff{1}{r} \fff{d f(r)}{d r}\right] + \nonumber \\ +
\fff{{g^*}^2}{2\mu r^2} \fff{1}{\Theta(\theta)}
\fff{d^2\Theta(\theta)}{d\theta^2} + E - V(r) = 0,
\end{eqnarray}
or, equivalently,
\begin{eqnarray}
\fff{1}{\Theta(\theta)} \fff{d^2\Theta(\theta)}{d\theta^2} = -
\fff{r^2}{f(r)}\left[\fff{d^2f(r)}{d r^2} + \fff{1}{r}\fff{d
f(r)}{d r}\right] - \nonumber \\ - \fff{2\mu
r^2}{{g^*}^2}[E-V(r)].
\end{eqnarray}

Since the term on the left hand side depends only on $\theta$ and
the term on the right hand side depends only on $r$, both terms
must be equal to a constant which we denote by $-\ell^2$.
Therefore, we obtain two ordinary differential equations:
\begin{equation}
\label{6} \Theta''(\theta) = -\ell^2\Theta(\theta),
\end{equation}
and
\begin{equation}
\label{7} f''(r) + \fff{1}{r} f'(r) + \left\{ - \fff{\ell^2}{r^2}
+ \fff{2\mu}{{g^*}^2}[E-V(r)]\right\}f(r) = 0.
\end{equation}

Equation (\ref{6}) has a solution
\begin{equation} \label{Theta}
\Theta(\theta) = e^{i \ell \theta}.
\end{equation}
If we assume the boundary condition
\begin{equation}
\Theta(0) = \Theta(2\pi),
\end{equation}
we end up with $|\ell| = 0, 1, 2, 3 \ldots$; $\ell$ is an integer.

Now we apply a rescaling in (\ref{7}) by
\begin{equation}
\rho = 2\beta r,\quad\quad \beta > 0,
\end{equation}
where $\beta^2 = - 2\mu E/{g^*}^2$. Furthermore we define
\begin{equation}
\label{gamma} n = \fff{\mu G Mm}{{g^*}^2\beta},
\end{equation}
and, by using $V(r)=-GMm/r$, equation (\ref{7}) becomes
\begin{equation}
\label{8} \tilde{f}'(\rho) + \fff{1}{\rho} \tilde{f}'(\rho) +
\left( - \fff{1}{4} - \fff{\ell^2}{\rho^2}+\fff{n}{\rho}\right)
\tilde{f}(\rho) = 0.
\end{equation}

If we consider $\tilde{f}(\rho) = \fff{1}{\sqrt{\rho}}u(\rho)$ in
(\ref{8}) we obtain
\begin{equation}
\label{Whittaker} u'' (\rho) + \left[-\fff{1}{4} + \fff{n}{\rho} -
\fff{(\ell^2 - 1/4)}{\rho^2}\right]u(\rho) = 0.
\end{equation}
For this equation, $\rho = 0$ is a regular singular point and
$\rho = \infty$ is an irregular singular point. Equation
(\ref{Whittaker}) is a confluent hypergeometric equation
\cite{Ince}, referred as Whittaker's equation \cite{PZ}. This
equation has a regular solution given by a hypergeometric series
which converges if, and only if,
\begin{equation}
n = |\ell| \pm \fff{1}{2} \pm k, \quad\quad k = 0, 1, 2, 3 \ldots
\end{equation}

By definition, $n \geq 0$. Assuming that the solution $u(\rho)$
satisfies the boundary condition
\begin{equation}
\ddd\lim_{\rho \to +\infty}u(\rho) = 0,
\end{equation}
and as $u(0)$ must be finite, we have
\begin{equation}
n = |\ell| + \fff{1}{2} + k, \quad\quad k =0, 1, 2, 3 \ldots
\end{equation}

So far we have the conditions
\begin{equation}
\begin{array}{lcl}
|\ell| & = &  0, 1,2,3 \ldots
\\
n &=& |\ell| + \fff{1}{2},\;\; |\ell| + \fff{3}{2},\;\; |\ell| +
\fff{5}{2} \ldots
\end{array}
\end{equation}
They can be regrouped in the form
\begin{equation}
\begin{array}{lcl}
n & = & \fff{1}{2}, \fff{3}{2}, \fff{5}{2}, \fff{7}{2} \ldots
\vtres\\ \ell & = & 0, 1, 2, \ldots, n - \fff{1}{2}.
\end{array}
\end{equation}

For each pair $n\ell$, the solutions $u_{n \ell}$ were obtained
using Maple V Release 4 Software  (Waterloo Maple Inc.). As long
as we obtain $u_{n \ell} (\rho)$, we may find the solution $f_{n
\ell}(r)$ of the original equation (\ref{7}). Once we know the
collection of functions $f_{n \ell}(r)$ and $ \Theta_{ \ell}(r)$,
their products generate the normalized solutions $\psi_{n \ell}$
of the Schr\"{o}dinger-type equation.

Now we define
\begin{equation}
P_{n\ell}(r)dr = \ddd\int^{2\pi}_{0} \psi^*_{n\ell} \psi_{n\ell} r
dr d\theta,
\end{equation}
which is equivalent to
\begin{equation}
P_{n\ell}(r)dr = r\left[f(r)\right]^2dr.
\end{equation}
As in the case of atomic wavefunctions, the mean radius is
obtained by
\begin{equation}
{r}_{n\ell} = \ddd\int^{\infty}_{0} r P_{n\ell}(r)dr =
\ddd\int^{\infty}_{0}\ddd\int^{2\pi}_{0}\left(\psi^*_{n\ell}r\psi_{n\ell}
\right)rdrd\theta,
\end{equation}
which is equivalent to
\begin{equation}
\label{10} {r}_{n\ell} = \ddd\int^{\infty}_{0}[rf(r)]^2dr.
\end{equation}

In this context, we now return to the rescaling factor $\beta$ and
consider the mean distance of Mercury as ${r}_{\frac{3}{2}0}$. (If
we associate Mercury with the first state ${r}_{\frac{1}{2}0}$,
all results are inconsistent with the observed mean planetary
distances. The orbits of Venus, Earth, Mars and Jupiter, for
example, definitely cannot be derived.) For $n=3/2$ and $l=0$, we
obtain for Eq. (\ref{Whittaker}) the solution
\begin{equation}
\fff{1}{\sqrt{\rho}}u(\rho) =
c_1\left(-e^{-\frac{1}{2}\rho}+e^{-\frac{1}{2}\rho}\rho\right),
\end{equation}
and so $f(r) = c_1 e^{-\beta r}(2\beta r - 1)$. Since
$\ddd\int^{\infty}_{0}r [f(r)]^2 dr= 1$ we have
\begin{equation}
c_1 = \raiz{\fff{4}{3}}\beta.
\end{equation}
From Eq. (\ref{10}) we then get
\begin{equation}
{r}_{\frac{3}{2}0} = \fff{7}{3} \; \fff{1}{\beta} = 0.387\; AU,
\end{equation}
where we have used the observed mean distance of Mercury. So, we
obtain
\begin{equation}
\beta = \beta_{\frac{3}{2}} = \fff{1}{0.387} \; \fff{7}{3} =
9.04/(3/2)\; AU^{-1}.
\end{equation}
Therefore, since $\beta$ is proportional to $1/n$ (see Eq.
(\ref{gamma})), for each value of $n$ we will consider
\begin{equation} \label{beta}
\beta_n=9.04\; (1/n)\; AU^{-1}.
\end{equation}

Its important to remark that the orbital radius of Mercury will be
the only input parameter used in this model.

\section{Results and discussions}

We denote by the triple $(n,\ell,r)$ the resulting numbers $n$,
$\ell$, and the corresponding mean planetary radius $r_{n \ell}$
(in AU), calculated from equation (\ref{10}). The first triple
found (1/2,0,0.055) does not correspond to any observed orbit in
our solar system. As previously mentioned, Nottale \cite{Nottale2}
also found this radius as the fundamental level in his
quantization law, and it was also obtained by Agnese and Festa as
the first state in their quantization condition \cite{AF1}. As
remarked in these two papers, the first extra-solar planet
discovered around the solar-type 51 Pegasus lies precisely at 0.05
AU from its star. Other subsequently discovered companions of
solar-type stars, which also fall around 0.05 AU, confirm this
result \cite{Schneider}.

The following two states (3/2,0,0.387) and (3/2,1,0.332) are
associated with the orbit of Mercury. The triples (5/2,0,1.05) and
(5/2,1,0.995) have very similar values for $r$ which are equal to
the mean planetary distance of Earth. The triple corresponding to
Venus is (5/2,2,0.83).

The set of triples (7/2,0,2.04), (7/2,1,1.99) and (7/2,2,1.82)
does not correspond to any observed planet in this region of the
solar system, but could be associated with the asteroid Hungarias,
at $1.94$ AU. This radius value was also predicted by Nottale in
his quantization law \cite{Nottale1} and nearly corresponds to the
radius found for a companion of the solar-type star 47 UMa B,
lying at 2.12 AU from this star (see \cite{Nottale2,AF1} and
references therein).

The state (7/2,3,1.54) clearly stands for the orbit of Mars. The
results obtained for $n=9/2$ are (9/2,0,3.37), (9/2,1,3.32),
(9/2,2,3.15), (9/2,3,2.88) and (9/2,4,2.49). They are in a very
good correspondence with the asteroid belt in the solar system.
The lower values 2.88 AU and 2.49 AU correspond to the mean
distances observed for the Ceres group (2.64 AU), the central peak
of the belt, and for Vesta (2.36 AU), which delimits its interior
ring. The upper values 3.32 AU and 3.37 AU correspond to the mean
distance observed for Camilla (3.48 AU), which delimits the
exterior ring. The intermediate value 3.15 AU has a precise
correspondence with the Hygeia group (3.16 AU), the main peak of
the asteroid belt.

For larger values of $n$, a great number of states is found.
Nevertheless (as pointed out by Nottale et al. \cite{Nottale3} to
explain the circularity of the orbits in the solar system), a
great number of states occupying the same region of space, most of
them with large eccentricities, leads to a strong chaos and to the
crossing of orbits, which on large time scales would generate the
condensation of states on the observed, approximately circular
orbits. Therefore, for $n>9/2$ we will consider only the states
with rotational symmetry, that is, with $l=0$ (see Eq.
(\ref{Theta})).

In this way, the states (11/2,0,5.03) and (15/2,0,9.34) can be
associated with the orbits of Jupiter and Saturn, respectively.
The states (17/2,0,12.0) and (19/2,0,15.0) give mean values around
the orbit of the asteroid Chiron, distant $13.7$ AU from the Sun.
The state (21/2,0,18.3) can be associated with the orbit of
Uranus, while the state (25/2,0,25.9) has a mean distance close to
the orbital radii of the already referred asteroids 1993 HA 2 and
1995 DW 2. The next state, (27/2,0,30.2), clearly stands for the
orbit of Neptune and, finally, the state (31/2,0,39.9) can easily
be associated with the orbit of Pluto.

As one can see, all the planets in the solar system (as well as
the asteroid belt) are fitted by the model. Apart the fundamental
radius $0.055$ AU (which should be ruled out for thermodynamical
reasons \cite{Nottale3}), we have only three empty orbits,
corresponding to the states (13/2,0,7.02), (23/2,0,21.9) and
(29/2,0,34.9). In what concerns the occupied states, the
deviations from the observed orbital radii are $15\%$ for Venus,
$10\%$ for Chiron, $6\%$ for Ceres and Vesta, $5\%$ for Uranus and
the asteroids HA2 and DW2, $3\%$ for Jupiter, Camilla and
Hungarias, $2\%$ for Saturn, $1\%$ for Mars and Pluto, and $0.5\%$
for Earth, Neptune and Hygeia.

Beyond the orbit of Pluto, we still have the states (33/2,0,45.2)
and (35/2,0,50.8), which could correspond to other mass peaks in
the Kuiper belt, a group of about 60 trans-Neptunian objects
orbiting up to 50 AU from the Sun \cite{Kuiper,Morbidelli}. [After
the completion of this work, it was announced the discovery of
Quaoar, the biggest asteroid discovered up to now in the Kuiper
belt \cite{NS}. Its stable orbit has mean radius around 42 AU,
which may correspond either to the state (31/2,0,39.9), associated
to Pluto's peak, or to the state (33/2,0,45.2). The relative
deviations are $5\%$ and $8\%$, respectively.]

\section{Conclusions}

The set of obtained results in calculating planetary orbits in the
present work, as well as those obtained by other authors, are
enough to encourage further studies. In all these theoretical
approaches, the existence of a fundamental radius around 0.05 AU
is predicted, and several planets have been recently discovered
orbiting at this radius in extra-solar systems. Furthermore, it
was predicted the existence of asteroids between Uranus and
Neptune \cite{Marcal}, a prediction confirmed by observations.

The Scr\"{o}dinger-type model presented in this work seems to
describe the observed mean distances from Mercury until the
asteroid belt. For the outer system we have considered only states
with rotational symmetry, obtaining good results as well, with
only three empty states. All the results were obtained with just
one input parameter, namely the orbital radius of Mercury.

The reader may be asking how a set of planets and asteroids with
so different masses can be described by just one equation. In
other words, why do we need only one input parameter (the value of
$\beta$ for Mercury), if equation (\ref{4}) depends explicitly on
the reduced mass $\mu$?

The answer may reside on the fact that our Schr\"odinger-type
equation is not a genuine quantum equation, but describes a kind
of diffusion process. In this context, the re-scaled ``Planck
constant" present in (\ref{4}) is related to the diffusion
coefficient ${\cal D}$ by $g^*=2m{\cal D}$ \cite{Nelson,Nottale3},
where $m$ is the mass under diffusion. Furthermore, for any planet
or asteroid gravitating around the Sun, the reduced mass $\mu$
coincides with its mass $m$. Therefore, remembering that $V(r)$
and $E$ are also proportional to $m$, it is easy to verify that
(\ref{4}), in fact, does not depend on $m$.

Another way of verifying this result is to look at equation
(\ref{gamma}). Using $\mu=m$ and $g^*=2m{\cal D}$, we obtain
$\beta = G M/(4{\cal D}^2 n)$. So, for a given diffusion
coefficient and a given stellar mass, the parameter $\beta$
depends just on $n$, as assumed in Eq. (\ref{beta}).

As a last remark, let us point out an observable difference
between the flat model presented here and spherical models. In the
approaches by Nottale {\it et al} \cite{Nottale3} and by Agnese
and Festa \cite{AF1}, there is a predicted orbit at $0.18$ AU from
the Sun, corresponding to $n=2$ (remember that, in their models,
Mercury corresponds to $n=3$). This orbit is also predicted
through the Bohr-like model presented in \cite{Marcal},
corresponding to the quantum numbers $n=1$, $n'=0$ (see Eq.
(\ref{marcal'})). As a little planet orbiting at the fundamental
radius $0.05$ AU is ruled out by thermodynamical constraints
\cite{Nottale3}, a body at $0.18$ AU would be the only new body
predicted to orbit near the Sun.

As we have seen in the previous section, such an orbit does not
exist in the present model. It is excluded by the flatness of our
solutions. Of course, this does not mean that it could not be
observed in extra-solar systems. Actually, orbits with radius
around this value were already observed \cite{Schneider}. This
fact could suggest that, in contrast to our solar system, the
corresponding planetary systems are not flat, which could indicate
that they were originated in a different way.

\section*{Acknowledgments}

We gratefully acknowledge the support and encouragement from
Kleber C. Mundim (UnB) and Marcelo M. Moret (UEFS). We would like
to thank Naelton Mendes de Araujo ( Sistema Telebras-Embratel) for
sending astronomical data about the discovery of the asteroids HA2
and DW2. Also Ricardo Fragelli (UnB) for the extraordinary support
on writing Maple programs for the solution of the
Schr\"{o}dinger-type equation. Finally ours thanks to Maristela
Rocha (UnB) and Hans-Peter Leeb for reading and providing helpful
suggestions on the manuscript. S.C. was partially supported by
CNPq, Brazil.


\begin{thebibliography}{}

\bibitem{Marcal} M. Oliveira Neto, Ci\^encia e Cultura (Journal of the
Brazilian Association for the Advancement of Science), 48 (1996)
166.

\bibitem{SC} S. Carneiro, Found. Phys. Lett., 11 (1998) 95.

\bibitem{Nottale1} L. Nottale, {\it Fractal space-time and microphysics:
towards a theory of scale relativity} (World Scientific,
Singapore), 1993, p. 311.

\bibitem{Nottale2} L. Nottale, A\&A, 315 (1996) L9.

\bibitem{Nottale3} L. Nottale, G. Schumacher, and J. Gay, A\&A, 322
(1997) 1018.

\bibitem{Nottale4} L. Nottale, G. Schumacher, and E.T. Lef\`evre,
A\&A 361 (2000) 379.

\bibitem{AF1} A. G. Agnese and R. Festa, Phys. Lett. A, 227 (1997) 165.

\bibitem{AF2} A. G. Agnese and R. Festa, Hadronic J., 21 (1998) 237.

\bibitem{AF3} A. G. Agnese and R. Festa, {\it Discretizing
u-Andromedae planetary system}, astro-ph/9910534, v2, 1999.

\bibitem{Arp} H. Arp, {\it Seeing Red: Redshifts, Cosmology and
Academic Science} (Apeiron, Montreal), 1998, p. 219.

\bibitem{MPC} Minor Planet Circular, 25441 (07/12/95) and
25184 (05/14/95), International Astronomical Union.

\bibitem{Schneider} J. Schneider, {\it Extra-solar planets
catalog}, \\ http://www.obspm.fr/encycl/catalog.html.

\bibitem{Greenberger} D. M. Greenberger, Found. Phys., 13
(1983) 903.

\bibitem{Nelson} E. Nelson, Phys. Rev., 150 (1966) 1079.

\bibitem{Naschie} M. S. El Naschie, E. Rossler, and I. Prigogine,
{\it Quantum Mechanics, Diffusion and Chaotic Fractals} (Pergamon
Press, Oxford), 1995.

\bibitem{Sidharth} B. G. Sidharth, Chaos, Solitons and Fractals,
15 (2003) 25.

\bibitem{Bergia} S. Bergia, F. Cannata, and A. Pasini, Phys. Lett.
A 137 (1989) 21, and references therein.

\bibitem{Hills} J. G. Hills, Nature, 225 (1970) 840.

\bibitem{HH} M. H\'{e}non and C. Heiles, Astrophys. J., 69 (1964)
73.

\bibitem{Laskar} J. Laskar, Nature, 338 (1989) 237.

\bibitem{SW} G. Sussman and J. Wisdom, Science, 257 (1992) 56.

\bibitem{Ince} E. L. Ince, {\it Ordinary Differential Equations}
(Dover Publications Inc., New York), 1956, pp. 184 and 464.

\bibitem{PZ} A. D. Polyanin and V. F. Zaitsev, {\it Handbook of
exact solutions for O.D.E.} (C.R.C. Press), p. 144.

\bibitem{Kuiper} G. P. Kuiper, in {\it Astrophysics: A Topical
Symposium}, ed. J. A. Hynek (McGraw-Hill), 1951, p. 357.

\bibitem{Morbidelli} A. Morbidelli, Celestial Mechanics and
Dynamical Astronomy, 72 (1999) 129.

\bibitem{NS} http://www.newscientist.com/news, 7 October 2002.

\end{thebibliography}
\end{document}